\documentclass[10pt,twocolumn]{article}
\usepackage{times}

\usepackage[utf8]{inputenc}
\usepackage{graphicx}
\usepackage{svg}
\usepackage{algorithm}
\usepackage{algorithmic}
\usepackage{subcaption} 
\usepackage[capitalize]{cleveref}
\usepackage{svg}
\graphicspath{ {images/} }

\title{Optimising Virtual Resource Mapping in Multi-Level NUMA Disaggregated Systems}
\author{Ewnetu Bayuh Lakew \and Petter Sv\"ard \and Erik Elmroth \and Johan Tordsson\\
Dept. of Computing Science, Ume{\aa} University, Sweden}

\begin{document}
\date{}
\maketitle

\begin{abstract}
Disaggregated systems have a novel architecture motivated by the requirements of resource intensive applications such as social networking, search, and in-memory databases. The total amount of resources such as memory and CPU cores is very large in such systems. However, the distributed topology of disaggregated server systems result in non-uniform access latency and performance, with both NUMA aspects inside each box, as well as additional access latency for remote resources. 
In this work, we study the effects complex NUMA topologies on application performance and propose a method for improved, NUMA-aware, mapping for virtualized environments running on disaggregated systems. Our mapping algorithm is based on pinning of virtual cores and/or migration of memory across a disaggregated system and takes into account application performance, resource contention, and utilization. The proposed method is evaluated on a 288 cores and around 1TB memory system, composed of six disaggregated commodity servers,  through a combination of benchmarks and real applications such as memory intensive graph databases. Our evaluation demonstrates significant improvement over the vanilla resource mapping methods. Overall, the mapping algorithm is able to improve performance by significant magnitude  compared the default Linux scheduler used  in system.

\end{abstract}

\section{Introduction}
\label{sec:intro}
Modern applications such as search, mail, social networking, online maps, and machine translation generate an extraordinary amount of data. To work upon these large data sets, the  applications need large amounts of scalable computing resources, often beyond the capability of a single physical server \cite{svard2014hecatonchire}. To provide resources for these critical and highly resource demanding workloads, an emerging paradigm of disaggregated systems are evolving. These systems enable cloud infrastructures to lift the physical limitations of a server where the application is deployed and allow aggregation and manipulation of resources more freely, for example by collating resources from multiple physical machines to an application. As resources become more available and as data centre infrastructure move from hardware-defined to software-defined infrastructures, resource disaggregation becomes a perfect fit for configuring and aggregating resources in pools, thus bringing greater flexibility and cost reduction. This new architecture brings many benefits to cloud systems such as elimination of current technological barriers of actual fluidity and scalability of cloud resources, reduction in the infrastructure and administration costs as well as drastic improvement in resource utilization \cite{goumas2017acticloud}.   

This opportunity, however, brings new challenges. Huge performance variability  can be observed depending on how the  virtual resources are aggregated and  mapped onto the physical resources for an application in such systems. The resource mapping should not only reduce, of totally avoid, contention for shared resources, it should also consider and optimize locality and proximity of resources composing the virtual machine (VM) where the application runs. 
For example, Performance optimization for locality can be achieved by making  each thread accesses local off-chip memory upon cache misses \cite{su2012critical}. However, maximizing memory locality  can create contention on  cache hierarchy by placing too many threads on the same Non-uniform Memory Access (NUMA) node. Performance limitation due to resource contention or remote memory accesses may limit application scalability \cite{terboven2008data}. Moreover, as multiple servers become one, and the possibility of collocating   many applications and manifestation of emergent behaviour increases. As a result performance of an application may vary   depending on which applications  run on the same  cache, NUMA-node, or  one of disaggregated servers.

Several previous studies have addressed the issue of contention in the memory hierarchy \cite{xu2010cache, majo2011memory}. While some of these works were designed by Uniform Memory Access (UMA) systems, others do not consider the impact of remote memory access on performance.
On the other hand, previous NUMA-aware works \cite{Blagodurov2010, schopp2012dynamic} for optimizing locality either consider small scale systems, i.e., a server with few NUMA nodes or use simulation to estimate the performance of a system in large scale disaggregated system. 
In this contribution we study the effect of core allocation in a large shared memory system when running workloads such as in-memory databases and benchmarks in a virtualized environment, on VM sizes ranging from small to larger than a physical server. We benchmark the system with core distributions ranging from best-case to worst-case scenarios on a disaggregated setup consisting of 6 commodity servers running the NumaConnect~\cite{numascaleWP} software and hardware. The workloads are classified into three different classes based on the extent of interference on other collocated applications. We next develop an online algorithm that decides the optimal virtual resource to physical resource configuration for all running applications using the classification, resource proximity and runtime hardware performance counters. This work is, as far as we know, the first in-depth measurement study on real dissaggregated hardware that takes into account  resource contention,  locality, and degree of interference to achieve good performance for applications.

\section{Background}
\label{sec:bg}
The gradual switch from single to multicore systems was in practice forced upon the community. After being true fore more than 40 years, Moores law finally approached end of life~\cite{simonite2016moore}, where making smaller transistors no longer increased performance as the limits of physics started to come into play at an atomic level. This means that as we can no longer count on single core performance to steadily increase, we have to use more cores instead to improve performance. While this approach has proved successful in many ways, it is hampered by a number of problems such as more complex programming and scaling models. A good understanding of how these systems work is vital in achieving maximal performance from the hardware. 
\subsection{The NumaConnect platform}
\begin{figure}[htp]
\centering
\includegraphics[width=6cm]{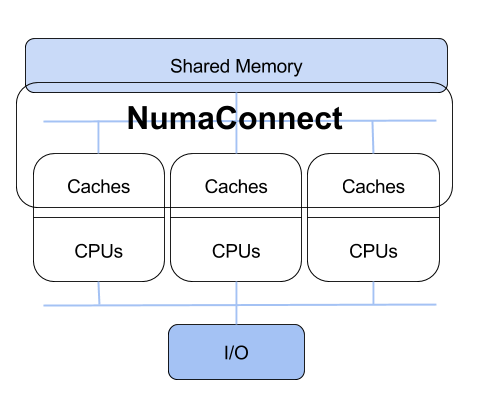}
\caption{Conceptual NumaConnect architecture}
\label{fig:NC}
\end{figure}

NumaConnect is a technology that allows access to hardware resources across nodes in a rack transparently. Its base is the NumaChip~\cite{numascaleWP} which incorporates shared memory control logic and a 7-way switch to connect to other NumaChip nodes, which means that all processors can share all memory and I/O devices through a single operating system instance. Each NumaChip contains an on-chip switch to connect to other servers in a NumaChip based system, eliminating the need to use a centralized switch. It is designed to communicate with the CPUs on the motherboard using the processor bus on a commodity server and bridge servers in a Cache Coherent Shared Memory manner. The system architecture is as follows. The NumaChip occupies one socket on the Motherboard, a NumaConnect network adapter bridges the communication on the Motherboard to other servers and NumaConnect cables connect the servers together. The end result is one unified, shared memory system of up to 4096 servers, acting as one cache coherent system without the need for a virtualization layer, or any centralized communication protocol. From an applications point of view, the system behaves as a single large node. An illustration of the concept can be seen in \cref{fig:NC}.

\label{sec:nscale}

The biggest difference between NumaConnect and other high-speed interconnect technologies is the shared memory and cache coherency mechanisms. These features allow programs to access any memory location and any memory mapped I/O device in a multiprocessor system with high degree of efficiency. It provides scalable systems with a unified programming model that remains the same from the small multicore machines used in laptops and desktops to the largest imaginable single system image machines that may contain thousands of processors. 

The advantages of such an architecture are:
\begin{itemize}
\item Any processor can access any data location through direct load and store operations, which leads to easier programming with usually less code to write and debug compared to distributed systems that support message passing only (e.g., MPI) programming models. 
\item Compilers can automatically exploit loop level parallelism that enables higher efficiency with less human effort.
\item Easier system administration of a unified system, as opposed to a large number of separate VMs in a cluster, requiring less effort.
\item Resources can be mapped and used by any processor in the system giving optimal use of resources in a single image operating system environment.
\end{itemize}

The fabric routing is controlled through routing tables that are initialized from system software, the NumaConnect Bootloader, at the BIOS level. The NumaConnect Bootloader ensures that all resources are available, unifies the resources in software, and presents all servers to the Linux kernel as one single system. When the operating system boots, it sees all the servers as one platform.

\subsection{Workload classification}
Applications behave differently and utilize the hardware in different ways depending on their characteristics. In the scope of this work we use a subset of the Animal Classes~\cite{xie2008dynamic}, as proposed by Xie et al. The Animal Classification scheme divides applications into four classes depending on their use of share last-level cache. In the scope of this work, we are not using the "Turtle" class but restrict ourselves to following three classes:

\begin{itemize}
    \item \textbf{Sheep:} Gentle and tame applications that are not significantly affected by sharing cache with other applications
    \item \textbf{Rabbit:} Fast and delicate applications whose performance rapidly degrade with insufficient cache allocation or sharing cache with other applications. 
    \item \textbf{(Tasmanian) Devil:} These applications access the cache very frequently, with very high miss rates. As a result, they do not benefit much from caching and tend to negatively impact other applications as they "thrash" the cache.
\end{itemize}

In this work, we assume that the applications have been classified by means of the animal scheme. The classification can be done by the application owner, by running it in a controlled environment, or it may already be known. In addition to the animal classes, we also classify how sensitive an application is to remote memory. This classification is rather coarse, an application is either sensitive or insensitive. This means that an application can be for example a "sensitive Rabbit" or an "insensitive Sheep". In scenarios where memory is scarce, sensitive applications are given more of the local memory on the nodes. In the scope of this work, application classification is static and does not change.

\section{Motivating Study}
\label{sec:motivation}
\begin{figure}[htp]
\centering
    \includegraphics[width=8cm]{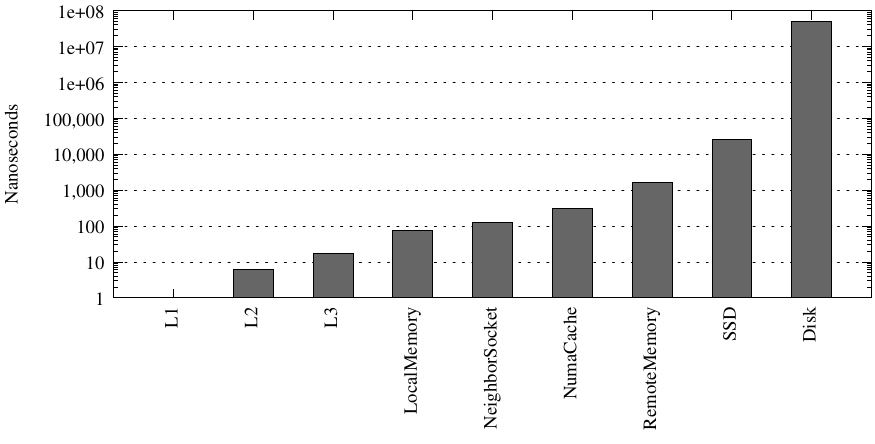}
    \caption{Latencies in the memory hierarchy}
\label{fig:numa_lat}
\end{figure}

The relative performance of an application on a given hardware platform is normally affected mainly by contention of shared resources, and in multicore environments, especially caches~\cite{zhuravlev2010addressing}. If several applications are updating the same cache, they might overwrite each others data, leading to cache contention. This holds for all cache levels, from L1 to LLC. In a NUMA environment, more applications share the same LLC than share L1, on the other hand the LLC is usually much larger~\cite{Jaleel:2010:HPC:1816038.1815971}, and less frequently updated. The NUMA topology also affects performance as accessing memory on another NUMA node comes with a higher latency due to a larger NUMA distance. In a disaggregated environment the NUMA topology is more complex with a much larger span of NUMA distances due to remote memory access. This factor is sometimes know as resource composition distance and affects relative performance because it requires network usage, which is orders of magnitude slower than the internal data buses. The difference in access latency between locations in the memory hierarchy can be significant, as illustrated in \cref{fig:numa_lat}.    

\subsection{Hardware setup}
\label{sec:setup}
The NumaConnect system used in both the motivating study and the main evaluation is a 6 node system with the following setup. 
\begin{itemize}
    \item 6 IBM x3755 M3 servers 
        \begin{itemize}
        \item 2x AMD 6380 (48 cores)
        \item 192 GB RAM
        \item 1 TB HDD
        \end{itemize}
    \item 6 Numascale NumaConnect N323 network adapters
    \item 1 Numascale Management Applicance
\end{itemize}

\begin{figure}[htp]
\centering
\includegraphics[width=5cm]{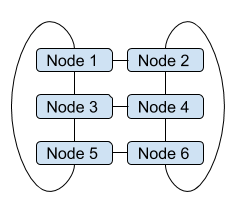}
\caption{Network Topology}
\label{fig:NT}
\end{figure}

The system has a total of 288 cores and 1176 GB of RAM. It is connected in a 2-dimensional torus fashion, as illustrated in \cref{fig:NT}, which means that the distance between nodes is never more than two hops. \cref{table:hardware} provides more information about the system.

\begin{table*}[h]
\centering
\begin{tabular}{|p{2.6cm}|p{9cm}|}
                   & AMD OpteronTM 6380                             \\ \hline
CPU(s)        & 288                                            \\ 
Thread(s) per core & 2                                              \\ 
Core(s) per socket & 8                                              \\ 
Socket(s)          & 18                                             \\ 
NUMA node(s)       & 36                                             \\ 
L1 cache           & 16K DCache, 64K ICache                         \\ 
L2 cache           & 2048K unified, shared by 2 threads in a core \\ 
L3 cache           & 6144K unified, shared by 8 cores               \\ 
                   & Numascale NumaConnect N323            \\ 
\end{tabular}
\caption{Hardware information}
\label{table:hardware}
\end{table*}

\subsection{Performance study}
\label{sec:study}
\begin{figure}[htp]
\centering
    \includegraphics[width=8cm]{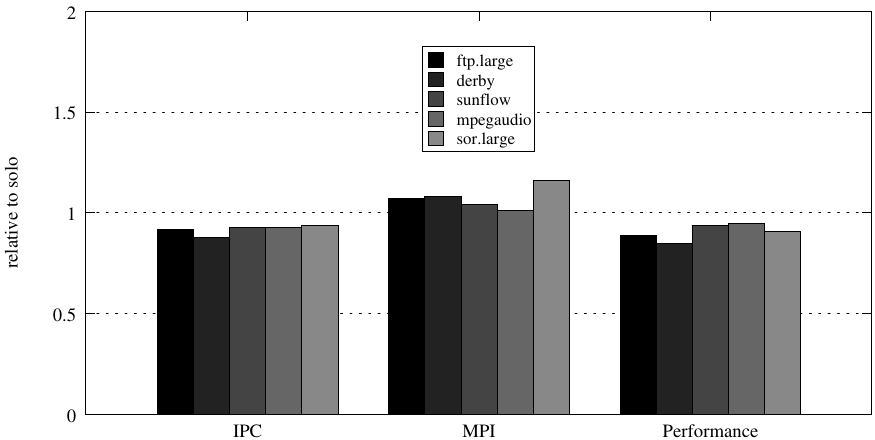}
    \caption{Neo4j}
\label{fig:neo4j}
\end{figure}

\begin{figure}[htp]
\label{fig:sockshop}
\centering
    \includegraphics[width=8cm]{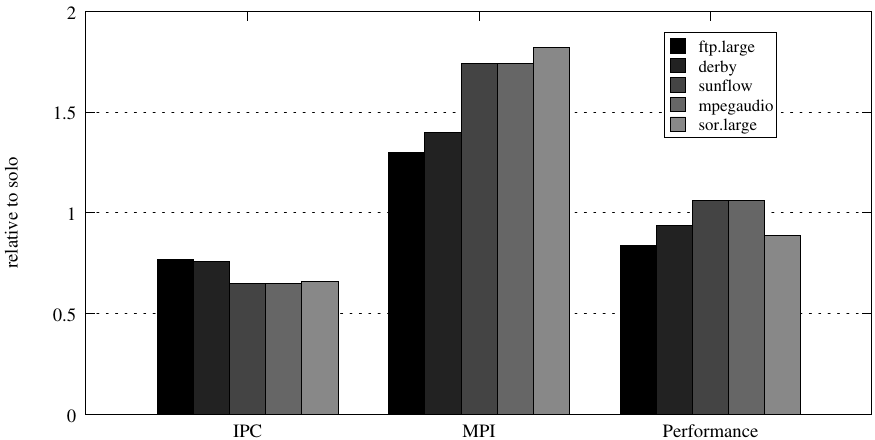}
    \caption{Sockshop}
\end{figure}

\begin{figure}[htp]
\label{fig:derby}
\centering
    \includegraphics[width=8cm]{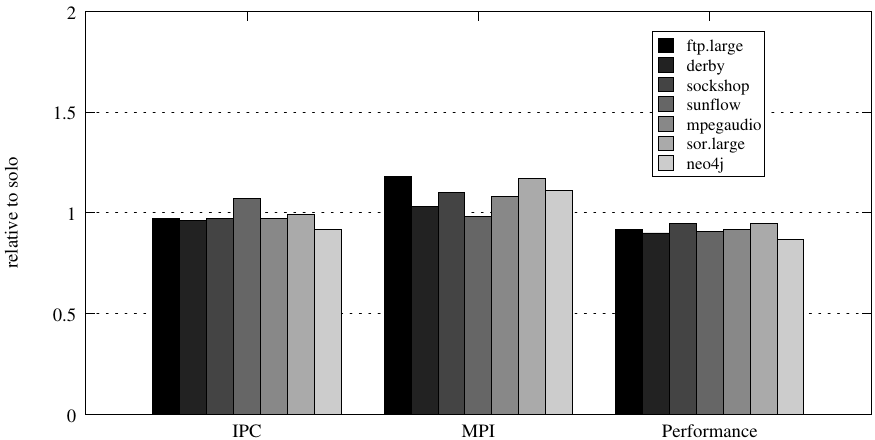}
    \caption{Derby}
\end{figure}

\begin{figure}[htp]
\label{fig:fft.large}
\centering
    \includegraphics[width=8cm]{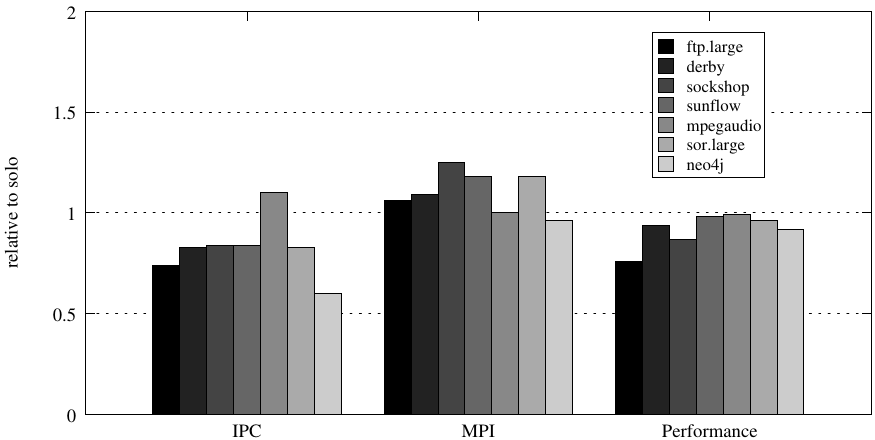}
    \caption{fft.large}
\end{figure}

\begin{figure}[htp]
\label{fig:sor.large}
\centering
    \includegraphics[width=8cm]{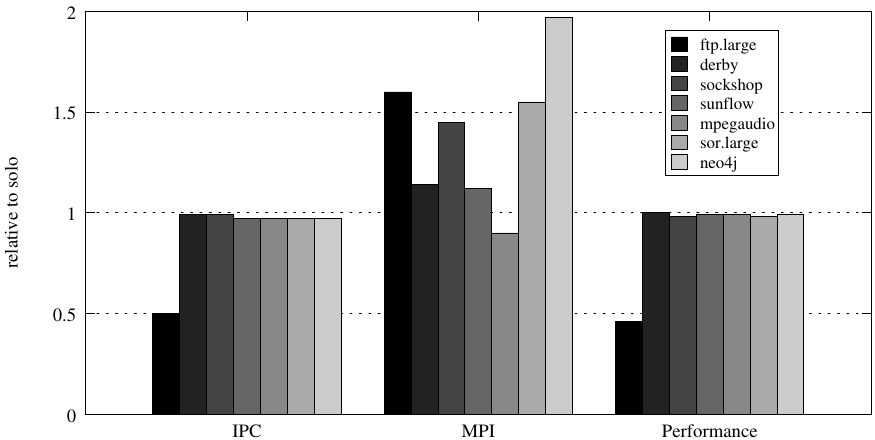}
    \caption{sor.large}
\end{figure}

\begin{figure}[htp]
\label{fig:mpegaudio}
\centering
    \includegraphics[width=8cm]{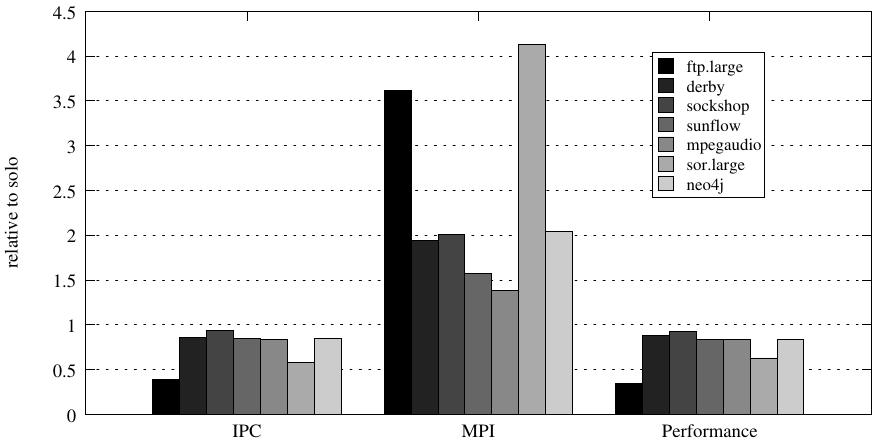}
    \caption{Mpegaudio}
\end{figure}

\begin{figure}[htp]
\centering
    \includegraphics[width=8cm]{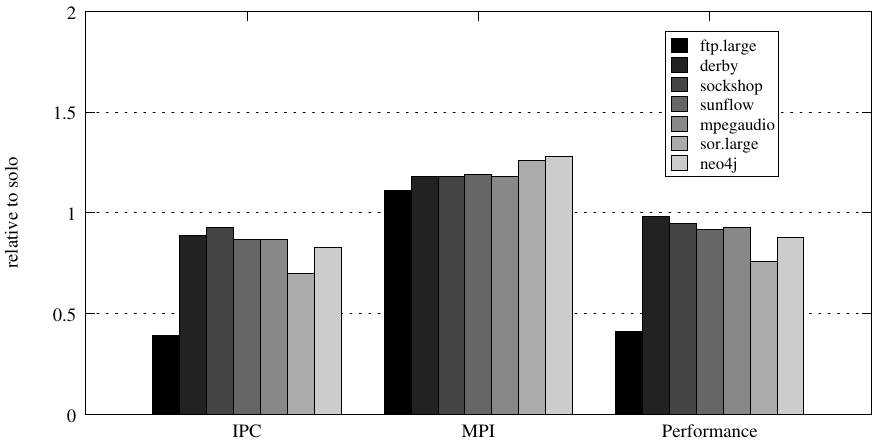}
    \caption{Sunflow}
\label{fig:sunflow}
\end{figure}

In order to investigate the impact of co-locating applications on the same NUMA node, we first ran a number of applications, showed in Table~\ref{table:applications}, solo and measured their Instructions-per-Cycle (IPC) and Misses-Per-Instructions (MPI). 
We then co-located another application with the test candidate on the same NUMA node and memory controller, so that they shared LLC, and ran the experiment again. The process was repeated 3-5 times and an average performance value relative to running solo was calculated. The MPI, IPC and performance relative to the solo run is presented in \cref{fig:neo4j} - \cref{fig:sunflow}. In this process, we also identified which of the animal classes the application belongs to, as shown in \cref{table:applications}.

\begin{table*}[h]
\centering

\begin{tabular}{l|@{}c@{}|@{}c@{}|@{}c@{}|@{}c@{}|@{}c@{}|@{}c@{}|@{}c@{}}
      & \textbf{Neo4j} & \textbf{Sockshop} & \textbf{Derby} & \textbf{\,fft\,} & \textbf{\,sor\,} & \textbf{\,mpegaudio\,} & \textbf{Sunflow} \\
    \hline
\textbf{Type} &  Database  &  \,Microservice\, & \,Benchmark\, & Benchmark & Benchmark & Benchmark & \,Benchmark \\
\textbf{Class} & Sheep & Sheep & Sheep & Devil & Devil & Rabbit & Rabbit   
\end{tabular}
\caption{Applications}
\label{table:applications}
\end{table*}

\subsection{Impact of NUMA distance}
\begin{figure}[htp]
\centering
    \includegraphics[width=8cm]{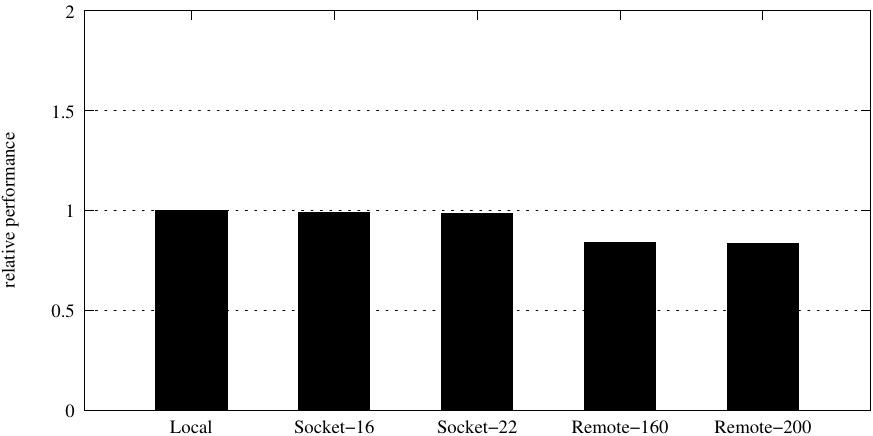}
    \caption{Relative performance of mpegaudio application under different NUMA-node distance.}
\label{fig:distance}
\end{figure}

Threads of a given application can be placed on the same NUMA nodes (local access) or different nodes (remote access). The remote access can have a great effect on performance due to variation in a way the nodes are connected. The NUMA distance in the system is 10 for local access, 16 or 22 for neighbor access, and 160 or 200 for remote access. To show the impact of connectivity on performance we ran the mpegaudio application with the same number of threads and nodes. The difference is which nodes are assigned, i.e., the connectivity of the selected nodes. \cref{fig:distance} shows the relative performance of the application  when assigned to different nodes, compared to running them on local nodes. As shown in the figure, performance of an application can be significantly affected depending the proximity of the node2. For example, in the mpegaudio case performance can drop as high as 17\% due to locality.  A well-designed system should take this effect into account and select the best-performing connectivity  for  those  applications that are access-sensitive. 

\subsection{Performance Indicators}
In such large system, co-located applications are expected to manifest emergent behaviors that lead to unhealthy resource contentions and undesirable performance interference. In order to detect and resolve undesirable performance behavior different Key Performance Indicators (KPI) such as application-related performance indicators, e.g., response time or throughput, and hardware performance counters, such as IPC and MPI~\cite{Zhang2013cpi} can be used. Application-related indicators may require modification of the application, making them intrusive to the application. Moreover, batch-oriented applications tend to provide this information once the execution of the application is completed, which is of no use for improving performance of the application during runtime. In this work, we use hardware performance counters such as IPC and MPI to resolve performance deviations.

\subsubsection{IPC as performance indicator}
 Instructions Per Cycle, IPC, is used to indicate an applications relative performance. IPC has been used for this purpose since the dawn of microprocessors and it is calculated by running a piece of code while calculating the number of machine instructions to complete it. High performance timers are then used to calculate the number of clock cycles used and the IPC value is given by dividing the number of instructions with the measured clock cycles. For this purpose, we utilize the Linux Perf tools~\cite{de2010new}. In general, relatively higher value of IPC for a given application implies better performance.

\subsubsection{MPI as performance indicator}
Misses Per Instruction, MPI, measures  cache-misses per instruction for different cache-levels and may signal interference among collocated applications due to resource contention~\cite{Iyer2009VMM}. In general, relatively lower value of MPI for a given application implies better performance.

\section{Resource Mapping}
\label{sec:method}
In this section, we describe our proposed mapping algorithm, used in the evaluation.

\subsection{Algorithm}
As the Numascale system behaves as one big machine the scheduling problem consists of assigning vcpus to physical cores, a process which we henceforth refer to as \textbf{mapping}. We do not use overbooking for cores, each core is running 0-1 vCPUs. If the system is at maximum capacity, we assume that a higher level of control will stop new arrivals to the system and possibly evict applications if needed.

The proposed algorithm works in two stages. First, remoteness is addressed, that is applications larger than a single node and thus requiring remote memory are mapped. This is done when applications arrive in the system ( See lines 2--11 under \cref{alg:alg1}). An application should be "sliced" as little as possible, that is spread out over as few servers as possible. However, if an application is using a lot of RAM but fever vcpu than a single node, some of the memory on that node is reserved to run other, smaller, VMs on the remaining cores.

\begin{table}[h]
\centering
\begin{tabular}{l|c|c|c}
       & Sheep & Rabbit & Devil \\
       \hline
Sheep  & X      & X     & X     \\
Rabbit & X      & -     & -     \\
Devil  & X      & -     & X   
\end{tabular}
\caption{Class Matrix}
\label{table:algo1}
\end{table}

In the second stage, the algorithm tries to place the applications as to cause as little interference between apps as possible (See lines 12--29 of \cref{alg:alg1}). This means that, for example, devil applications are not placed with rabbits or other devils. The full matrix can be seen in Table~\ref{table:algo1}.

\begin{table}[h]
\centering
\begin{tabular}{l|c|c|c}
       & Sheep & Rabbit & Devil \\
       \hline
Socket  & 1      & 4     & 7     \\
Numa Node & 1      & 5     & 8     \\
Server Node  & 1      & 6     & 9
\end{tabular}
\caption{Benefit Matrix}
\label{table:algo2}
\end{table}

When an application arrives, it is first placed according to these classifications. If the system is nearing its capacity and a good placement is not possible, we consider adjusting the placements on the whole system. The performance of the applications on the system are then monitored by means of either IPC or MPI depending on which  metric is selected (See line 15 of \cref{alg:alg1}). If performance drops, we analyze IPC values or cache hit/miss rates to try to identify if any applications are interfering with each other. In this case they need to be separated. In order to better predict the benefit of moving an application we setup a table with values 1-10 for each class of applications how much they would benefit from moving to their own socket, numa node or server node. This table, illustrated in Table~\ref{table:algo2}, is dynamically updated during runtime and, hence, the algorithm can make better mapping decisions over time.

\renewcommand{\algorithmicrequire}{\textbf{Configuration parameters:}}

\begin{algorithm*}[!t]
\caption{High Level Description of the Mapping Algorithm.}
\label{alg:controller}
{
\begin{algorithmic}[1]

\REQUIRE{
    $VMs$, list of running VMs \\
	$\bar{p_i}$ expected performance for $VM_i$, $p_i$, measured performance for $VM_i$ \\
	
	$c_i$, the interference class where $VM_i$ belongs to  \\
	$a_i$, policy regarding affinity for  $VM_i$   \\
	$R$, the hardware resource layout showing cores and memory arrangement\\
	$N_i$, neighbour VMs for $VM_i$ 
	$duration$, decision interval for the algorithm \\
	$T$, the maximum threshold value where relative performance deviation is tolerated  \\
	$A_{vms}$ list of VMs whose performance is negatively affected

}

\medskip

\WHILE{true}
\IF { $VM_i$ is a new arrival}
    \IF {Free slot is good to $VM_i$  the  given $c_i$, $a_i$}
        \STATE Map $VM_i$ 
        \STATE Update $VMs$
     
    \ELSE
        \STATE Choose which running VMs and where to reshuffle   to get suitable free slot  $VM_i$
        \STATE Remap the selected  VMs to their new configurations
        \STATE Map $VM_i$ 
        \STATE Update $VMs$
     \ENDIF
  \ELSE
\STATE Initialize $A_{vms}$ =\{\}
\FOR{ each $VM_i$ }
 
     \IF {$(\bar{p_{i}} - p_{i} )/ \bar{p_{i}} \geq T$}
       \STATE Add $VM_i$ to affected $A_{vms}$
      \ENDIF
 \ENDFOR

   \IF {$A_{vms}$ is not empty }
    \STATE Sort $A_{vms}$ by  $\bar{p_{i}} - p_{i} )/ \bar{p_{i}}$ 
    \FOR{ each $A_{vms}$ }
        \STATE Build potential neighbor list, $N_i$, based on $c_i$, $a_i$ for $VM_i$
        \STATE Compute new configuration  for $A_{vms}$  using neighborhood list and resource layout, $R$ that has least reshuffle
        \IF{current configuration is different from previous on for $VM_i$}
            \STATE Remap $VM_i$ to the new configuration 
            \STATE update Benefit Matrix
         \ENDIF
    \ENDFOR

    \ENDIF

\ENDIF
\STATE sleep for $duration$
\ENDWHILE

\end{algorithmic}
}
\label{alg:alg1}
\end{algorithm*}

\section{Experimental Evaluation}
\label{sec:setup}
The experiments were conducted on hardware described under section \cref{sec:setup} running a Numascale-modified CentOS7 as host OS and KVM as a hypervisor.
Our algorithm, described in the previous section, was implemented in python together with functionality for monitoring and aggregating the hardware performance counters. The algorithm controls the virtualized instances through the Libvirt API. The source code is available on github upon request.

\subsection{Instance types and numbers}
To emulate a typical cloud environment as well as the added values of such hardware four VM types were used: \texttt{small}, \texttt{medium}, \texttt{large} and \texttt{huge}. The configuration for the four VM types is shown on \cref{table:VM_types}. As seen in the table the \texttt{huge} VM type goes beyond the physical server boundary (i.e. its size is one  and half physical servers) to show the capabilities of disaggregated infrastructure to host resource demanding applications while the rest are designed to emulate VM sizes that are typical in commercial cloud offerings. To load the system 12 small VMs, 4 medium VMs, 2 large VMs, and 2 huge VMs were hosted at the same time. All guests run the Ubuntu OS.

\begin{table}[]
\centering
\begin{tabular}{|l|l|l|}
\hline
\textbf{VM Type} & \textbf{Number of Cores} & \textbf{Memory (GB)}\\ \hline
Small            & 4                        & 16          \\ \hline
Medium           & 8                        & 32          \\ \hline
Large            & 16                       & 64          \\ \hline
Huge             & 72                       & 288         \\ \hline
\end{tabular}
\caption{VM types used for the experiment.}
\label{table:VM_types}
\end{table}

To investigate performance of the placement heuristics it is important to use a spectrum of applications with different performance characteristics and hardware utilization. This section discusses workload classification and describes which workloads we have selected, and why.

\subsection{Selected workloads}
For this evaluation we selected a mix of a real-world application, a demo application and synthetic benchmarks for the evaluation. The real world application is Neo4j~\cite{miller2013n4j}, which is a graph database. Neo4j is a both CPU and memory intensive application, implemented in java. To generate load, a load generator, LDBC Load Generator, provided by Neo inc was used. To investigate performance of cloud applications we use the Sockshop Microservices Demo~\cite{sockshop} together with a load generator that simulates users shopping for socks. Both these applications scale well with increased resource allocation. Finally, two synthetic benchmarks were used, SPECjvm2008~\cite{specjvm2008}, a benchmark suite for measuring the performance of a Java Runtime Environment (JRE), contains several real life applications and benchmarks, and  Stream~\cite{stream}, which measures memory bandwidth and  and the corresponding computation rate for simple vector kernels. All the workloads were configured  to scale according the VM type they run.

\subsection{Results}
\label{sec:emethod}
This section  presents analysis of the results from the runs using the default scheduling mechanisms in KVM and Linux (henceforth "Vanilla") as well as our shared memory-aware algorithm.
\subsubsection{VM Resource Composition}
\begin{figure}[htp]
\centering
    \includegraphics[width=8cm]{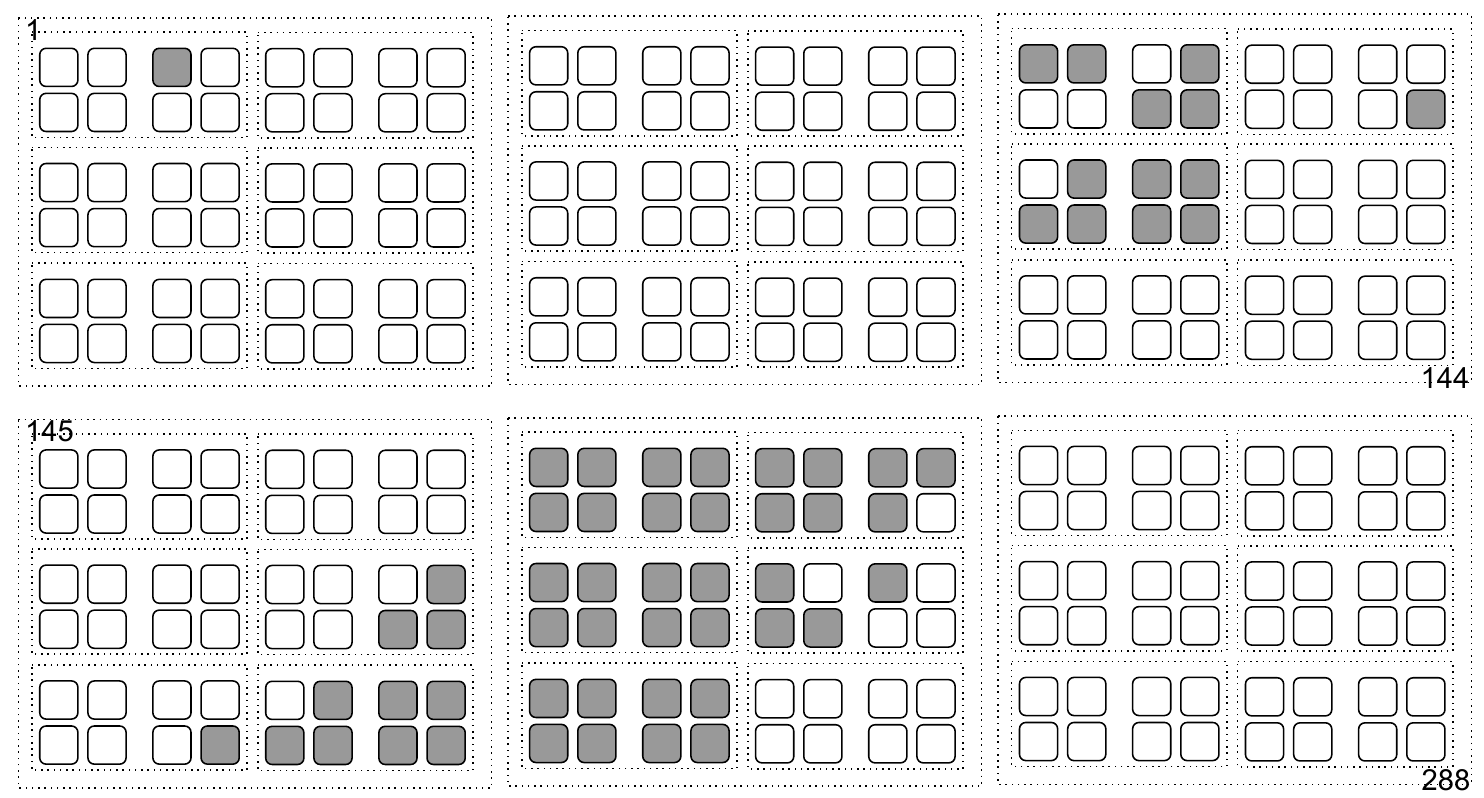}
    \caption{Core mapping, Huge VM, Vanilla algorithm}
\label{fig:map_vanilla}
\end{figure}

A close look at  the resource composition of a VM from the disaggregated resource pool under the two algorithms shows a clear difference. The reason behind and its implication on application performance is discussed in this section. 
\paragraph{Vanilla algorithm.}
Qemu/KVM is tightly integrated with Linux and each vcpu provided to the guest runs as a thread on the kernel~\cite{Habib2008}. This means that the vcpus are scheduled by the linux scheduler like any other threads. The linux scheduler was originally designed for single-core systems and as such, is not optimal for multi-core NUMA systems~\cite{Lozi2016}. This means that threads might not run as close to the memory they are working on as possible and for a shared memory system like the Numascale system used in this evaluation, the problems are exaggerated due to the performance penalty in accessing memory on another server. The scheduler, in its default state, also moves threads around which means that performance can vary greatly during a single run, and between runs. A snapshot of the vcpu mappings during a run is shown in \cref{fig:map_vanilla}, but as discussed, this mapping changes during runtime due to variations in load and the inner workings of the linux scheduler. Also note that some of the cores are overbooked. It is possible to tune the linux scheduler, for example using the compact scheme that tries to gather threads belonging to the same application or round robin-scheduling but as we have opted to control the mapping through libvirt we consider this to be outside the scope of this work.

\begin{figure}[htp]
\centering
    \includegraphics[width=8cm]{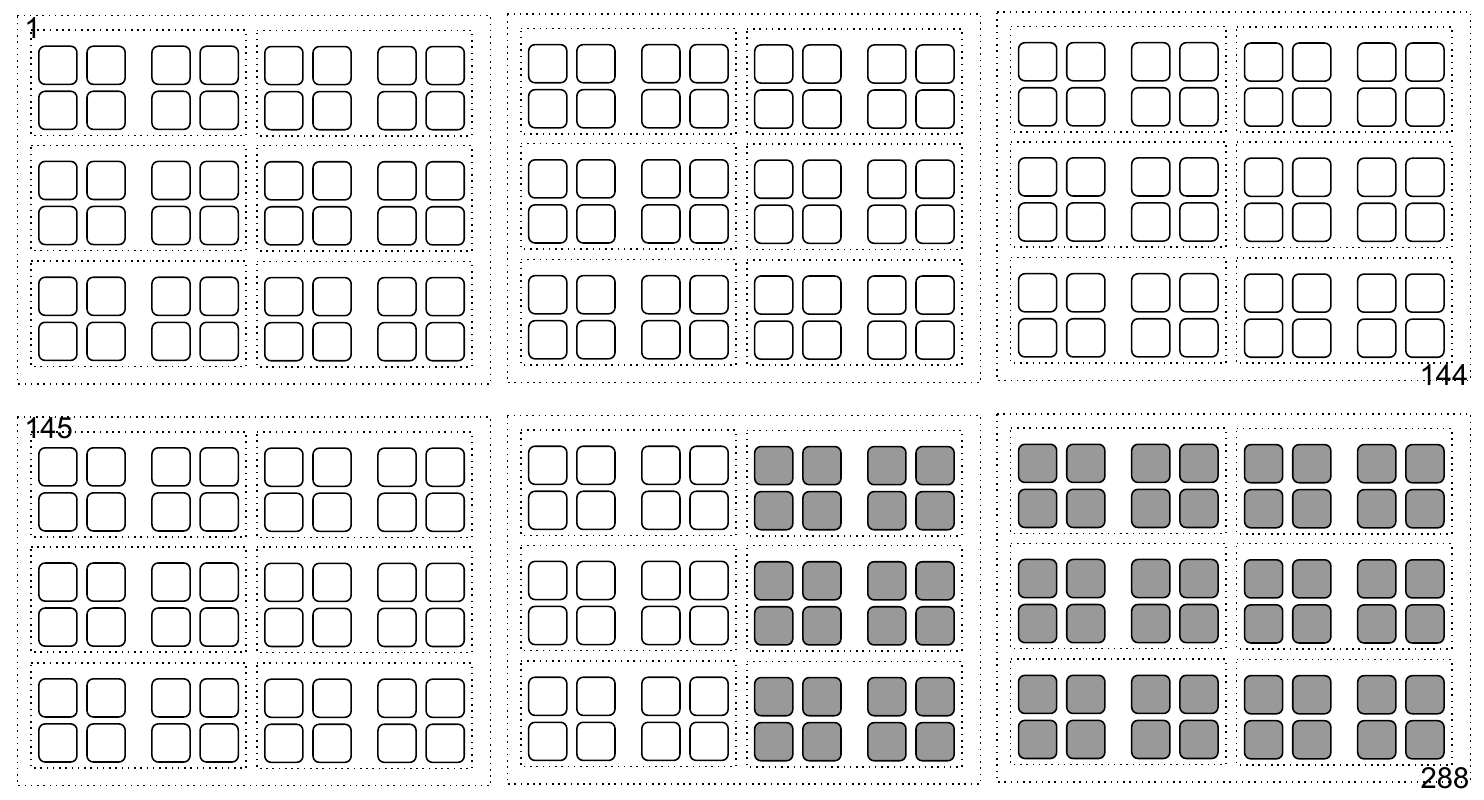}
    \caption{Core mapping, Huge VM, shared memory algorithm}
\label{fig:map_algorithm}
\end{figure}

\paragraph{Shared memory algorithm.}
As the proposed algorithm tries to keep cores close to memory and bring the cores of a VM together, the mapping is less random and does not vary over time in our experiment. The mapping of the same Huge VM as shown in the previous section using the proposed algorithm is shown in \cref{fig:map_algorithm}.

\subsubsection{Performance of different applications under the two algorithms}
This section presents how application performance is affected under the vanilla and shared memory algorithms. The shared memory algorithm  has two variants depending on the choice of  performance metric, \emph{IPC} or \emph{MPI}. To avoid confusion with performance metric, we will use \emph{SM-IPC} and \emph{SM-MPI} when referring shared memory using either IPC or MPI as a decision metric, respectively. All experiments were re-run three times and the results presented in the graphs are the average of the three runs.

\begin{figure}[!t]
\centering
    \includegraphics[width=8cm]{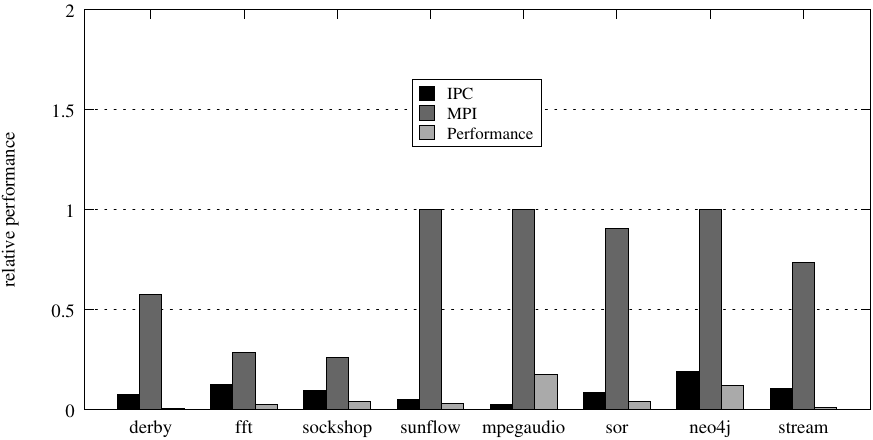}
    \caption{Relative performance (IPC, MPI, and application performance) for different applications using  vanilla algorithm.}
\label{fig:vanilla}
\end{figure}

\begin{figure}[!t]
\centering
    \includegraphics[width=8cm]{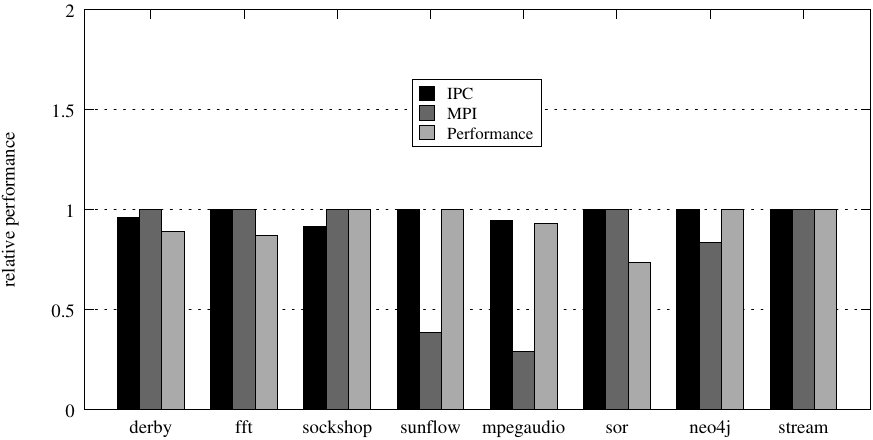}
    \caption{Relative performance (IPC, MPI, and application performance) for different applications using  SM-IPC.}
\label{fig:ipc}
\end{figure}

\begin{figure}[!t]
\centering
    \includegraphics[width=8cm]{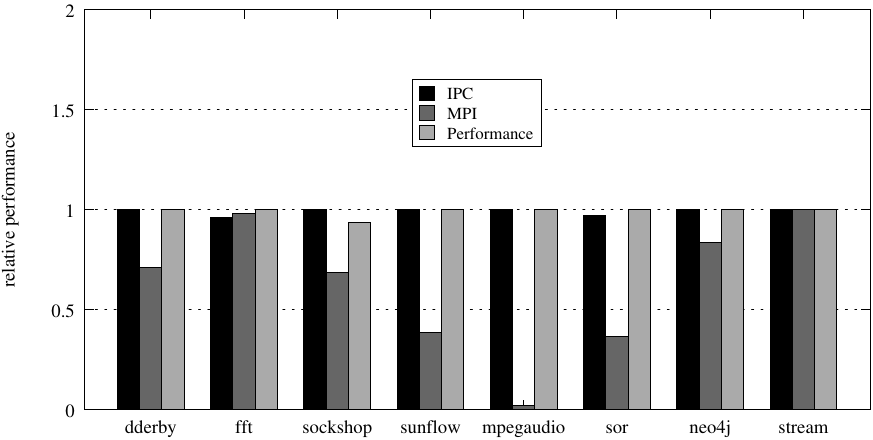}
    \caption{Relative performance (IPC, MPI, and application performance)  for different applications using  SM-MPI.}
\label{fig:mpi}
\end{figure}

\cref{fig:vanilla,fig:ipc,fig:mpi} present relative performance results  of different applications under \emph{vanilla}, \emph{SM-IPC} and \emph{SM-MPI} algorithms, respectively. The x-axis in the graphs shows the different applications, whereas the y-axis shows the  performance of each application relative to the three algorithms employed. For all applications  except Neo4j and Sockshop the results presented in the graphs are for \emph{medium} VM type. The VM type used for Neo4j is \emph{huge} while  for Sockshop \emph{small} is used.  Observed behavior of the algorithms under different VM type is presented under \cref{sec:vmtype}. From Figs., one can observe that for all applications when  \emph{vanilla} algorithm is employed, performance becomes  very poor while  \emph{SM-IPC} and \emph{SM-MPI}  provides better and comparable performance   for all  applications.

The results show that performance of an application significantly varies across different runs under \emph{vanilla} algorithm while the variation is negligible under both \emph{SH-IPC} and \emph{SH-MPI}. For example, the ratio between the standard deviation and average performance of the runs is above 0.4 for all applications under  \emph{vanilla} algorithm indicating its unpredictable behavior. On the other hand, such ratio remains below 0.04 for both \emph{SH-IPC} and \emph{SH-MPI}.  Moreover, the performance impact varies depending on the type of applications. For example, performance improves by 215x, 33x, 25x, 34x, 5x, 17x, 8x and 105x under \emph{SH-IPC} and 241x, 37x, 23x, 34x, 5x, 23x, 8x, and 105x  under \emph{SH-MPI}  compared to \emph{vanilla} for Derby, fft, Sockshop, Sunflow, mpegaudio, Sor, Neo4j, and Stream applications, respectively.  There are three contributing factors for such huge variation across applications. These are resource contention, overbooking and NUMA distance. The \emph{vanilla} algorithm is oblivious to application sensitivity to interference and distance. On top of that, sometimes some cores are overbooked making matters worse. The combined effect of the three factors with different degree of magnitude contributes to the the variability of the performance factor values shown above.  However, both \emph{SH-CPI} and \emph{SH-MPI} considers behavior of neighbourhood applications and memory proximity when mapping vcpus to physical cores while also ensuring  a given core is  allocated to a single application at time. Thus, both algorithms prevent resource overbooking while also continuously adapting the system to minimize the performance impacts of  resource contention and NUMA distance.  

\subsubsection{VM Size Matters}

\label{sec:vmtype}
\begin{figure}[!t]
\centering
    \includegraphics[width=8cm]{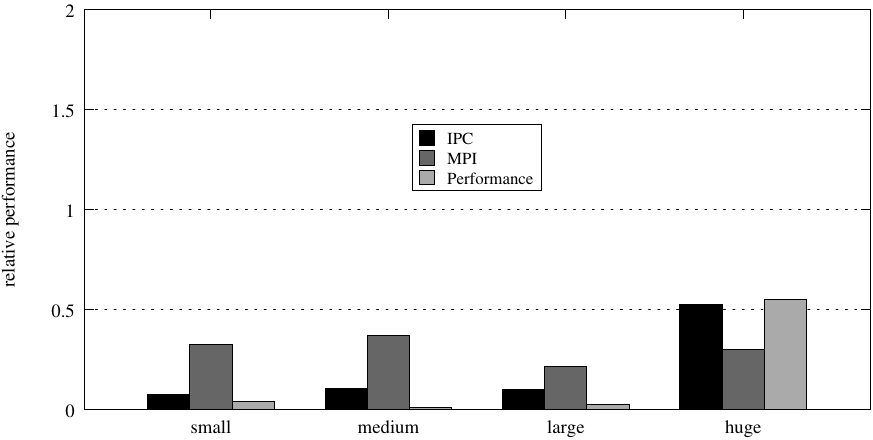}
    \caption{Stream: Relative performance (IPC, MPI, and application performance) for the four VM Types  under vanilla  algorithm.}
\label{fig:vm_type_vanilla}
\end{figure}

\begin{figure}[!t]
\centering
    \includegraphics[width=8cm]{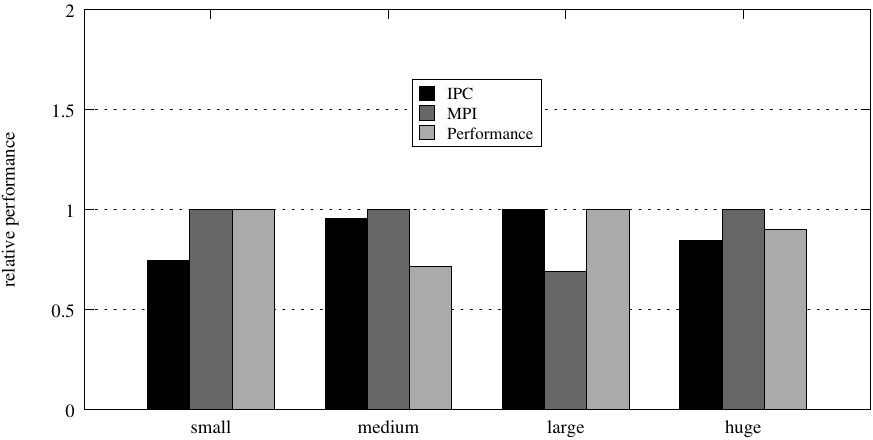}
    \caption{Stream: Relative performance (IPC, MPI, and application performance) for the four VM Types using   SM-IPC.}
\label{fig:vm_type_ipc}
\end{figure}

\begin{figure}[!t]
\centering
    \includegraphics[width=8cm]{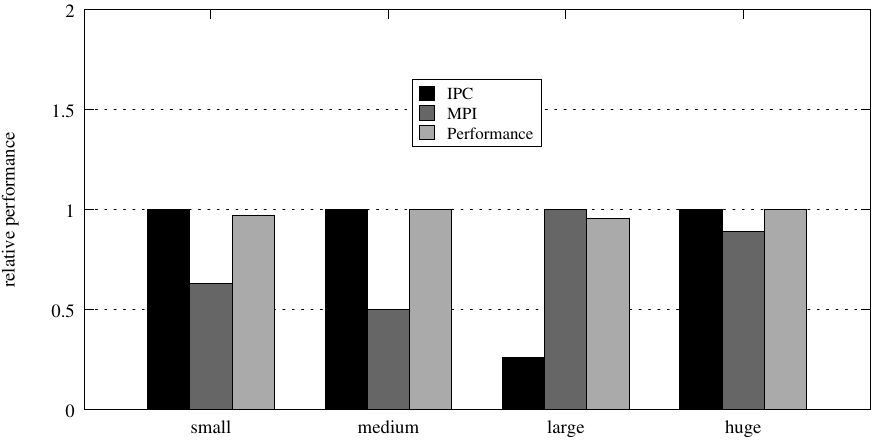}
    \caption{Stream: Relative performance (IPC, MPI, and application performance) for different applications using  SM-MPI.}
\label{fig:vm_type_mpi}
\end{figure}

This section presents whether VM size matters when applying \emph{vanilla}, \emph{SM-IPC}, or \emph{SH-MPI} algorithms when mapping virtual resource to physical resources  under disaggregated systems. Each experiment was re-run three times and the graphs show the averages of the three runs. \cref{fig:vm_type_vanilla,fig:vm_type_ipc,fig:vm_type_mpi} present relative performance results  of small, medium, large and huge VM types for  stream application under the \emph{vanilla}, \emph{SM-IPC} and \emph{SM-MPI} algorithms, respectively.
The results show that relative performance of stream application significantly varies depending on the VM size  under \emph{vanilla} algorithm while the variation is negligible under both \emph{SH-IPC} and \emph{SH-MPI}. Relative performance is 48x,
105x, 41x, 2x under \emph{SH-IPC} and 47x, 105x 39x, 
2x  under \emph{SH-MPI}  compared to \emph{vanilla} for small, medium, large, and huge VM sizes, respectively. Relative performance difference significantly drops for huge VM size. This can be  due to the fact that the chance of locality increase  as huge VM requires relatively higher number of cores. This can also result in less interference as the number of applications sharing NUMA nodes decreases. Moreover, relative performance variation across different runs decreases as the VM size increases for the \emph{vanilla} algorithm. On the other hand, relative performance variation remains insignificant under the \emph{SH-IPC} and \emph{SH-MPI} algorithms irrespective of VM size. Even though the performance variation across runs decreases, under \emph{vanilla}, it still remains significant. For example, under the \emph{vanilla} algorithm, the ratio between the standard deviation and average performance of the runs is above 0.2 for all VM sizes indicating its unpredictable behavior. Overall, the mapping algorithm improves performance of the mpegaudio by on average 50x compared the default Linux scheduler used in the system.


\section{Related Work}
Panagouirgious demonstrates in his thesis that memory location does impact performance in a NUMA system~\cite{panuourgias2011}. His evaluation uses specialized benchmarks on an isolated node. Lepers et al.~\cite{lepers2015thread} study the effects of asymmetric interconnect on a widely available x86 system and 
designed  a dynamic thread and memory placement algorithm that considers how the nodes are connected. 

Mayo and Gross~\cite{majo2011memory} propose a thread placement algorithm that reduces data locality to avoid the performance degradation caused by cache contention. Rao et al.~\cite{rao2013optimizing} propose a Bias Random vCPU Migration  algorithm
that dynamically migrates vCPUs to minimize the system-wide
uncore penalty. Tang et al.~\cite{tang2013optimizing}  investigates the impact of NUMA for several Google’s key web-service workloads in large-scale production  computers and focuses on the tradeoffs between memory access locality and the impact of cache contention.

While all the above works tried to address problem of locality and interference in NUMA systems, they are limited to a single server system. Our work is the first attempt in disaggregated system composed of commodity hardware.
\label{sec:rw}

\section{Conclusions}
\label{sec:concl}
This paper presents a mechanism that automatically composes resources from a pool of disaggregated commodity servers. Resources can be composed  for any type of applications without being limited by the physical size of traditional servers. A rigorous  performance analysis was performed  using different application to understand the performance implications of different configurations in  such systems. We proposed a mapping algorithm that enables infrastructure providers to use their resources more efficiently, better co-locate applications, and serve more users. Overall, the mapping algorithm is able to improve the
performance of an application on average up-to 50x  compared the default Linux scheduler used  in system.

In an extension of this work, we plan to study the effects of tuning the Linux scheduler to lessen the degree of randomness. Another interesting aspect is to use "memory follows cores" by using the memory migration technique available in libvirt. 

\label{sec:concl}


\end{document}